# Underwater Node Localization using Optoacoustic Signals

Muntasir Mahmud, Mohamed Younis, Gary Carter and Fow-Sen Choa
Department of Computer Science and Electrical Engineering,
University of Maryland Baltimore County
Baltimore, Maryland, USA
mmahmud1, younis, carter, choa@umbc.edu

*Abstract*—Localization of underwater networks is important in many military and civil applications. Because GPS receivers do not work below the water surface, traditional localization methods form a relative topology of underwater nodes (UWNs) and utilize either anchor nodes or floating gateways with dual transceivers in order to determine global coordinates. However, these methods introduce logistical complications and security risks in deploying the anchor and/or surface gateways. This paper tackles such an issue by proposing new localization techniques which can remotely localize UWNs using optoacoustic signals. In our approach, GPS coordinates are transmitted from air to the UWN via creating an underwater temporary isotropic acoustic transmitter with the optoacoustic process. We analyze the process of controlling the shape and size of the plasma to create the isotropic acoustic transmitter and experimentally validate the generation of isotropic acoustic signals. Then two methods of localization are proposed for static and dynamic UWNs. Finally, the simulation results with experimental values show the effectiveness of our approach. Comparing to the traditional techniques, our approach achieves the same accuracy without using any surface or underwater anchor nodes.

## I. Introduction

Recent years have witnessed significant advances in underwater networking technologies, motivated by applications such as search and rescue, security surveillance, sea-based combat, marine biology, etc. For these applications, underwater node localization is needed to establish and maintain a connected network topology. Moreover, localization is a vital requirement for the effective utilization of the sensed data by UWNs. However, an underwater network can easily get partitioned and some nodes can become unreachable because UWNs are generally mobile or drifted by the water current. Unlike terrestrial networks, GPS signal is not available underwater because it uses electromagnetic signals which have high attenuation in water medium. Therefore, an UWN cannot determine its own GPS coordinates like surface nodes.

Many localization methods are proposed to solve this problem. However, the conventional methods use underwater anchor based localization techniques. Instead of using a global coordinate system such as GPS, these methods establish a relative coordinate system in which the UWN positions are defined relative to one another. Global localization is possible by integrating surface (floating) nodes or using Dive and Rise (DNR) type anchor nodes. The floating node, e.g., a buoy or a boat, serves as a gateway for connecting the UWNs with satellites. Such a gateway is equipped with dual transceivers, one a radio for receiving GPS coordinates and another acoustic for transmitting that GPS coordinates to underwater nodes. A DNR anchor node rises above the surface to get GPS coordinates and while sinking, they broadcast their positions [1]. However, a floating gateway or DNR node requirement has several significant shortcomings, including the logistical constraint that complicates the deployment. Additionally, such deployment could expose the underwater network to security risks. For example, for military and security-sensitive applications, the gateway and DNR node could be located, and consequently, the presence of underwater nodes could be uncovered. Furthermore, it complicates the mobile underwater networks operation by imposing the need for fine-grained coordination during motion.

Avoidance of gateway and DNR nodes requires the development of a localization scheme with cross-medium communication technique. However, no single type of wireless signal can operate well across different mediums for long distances. For example, high-frequency radio waves can transfer data near light speed in the air but rapidly die after entering the water. Although low-frequency radio waves have a lower absorption coefficient in water, building antennas capable of radiating such long waves underwater is challenging. Visible light communication can be effective for short to moderate ranges, but the beams quickly get scattered and cannot support long-range communication [2]. Acoustics has been the preferred method of communication in the underwater environment [3]; however, an acoustic signal mostly attenuates when crossing the water surface.

This paper proposes a viable option for conducting UWN localization without the need of DNR or surface-based reference nodes. The idea is to employ the optoacoustic signals for establishing cross-medium communication links [4]. The optoacoustic effect refers to the generation of an acoustic signal when high-intensity light impinges on a liquid medium like water. This energy conversion process could be divided into two mechanisms, linear and nonlinear. The properties of the water medium do not change in the linear case. On the other hand, the physical properties of the water medium change in a nonlinear optoacoustic mechanism; specifically, water becomes vapor which creates cavitation bubbles [5]. We are leveraging the advantages of the nonlinear optoacoustic process for localization because it is suitable for reaching underwater receivers far from the surface. The Sound Pressure Level (SPL) of the nonlinear optoacoustic process is better than the linear counterpart. The simulation results in [5] have shown that the SPL for a linear optoacoustic process yields up to 140 dB re 1 µPa. Meanwhile, the SPL reported in [6] for a nonlinear optoacoustic effect is over 210 dB re µPa at 1 m. Therefore, Our method uses the nonlinear optoacoustic signal to achieve maximum localization coverage.

In this paper, we first discuss the process of generating optoacoustic signals and controlling the shape and size of the plasma to generate an isotropic acoustic signal. Then we experimentally verify such process by measuring the generated acoustic signal at $0^0$, $45^0$ and $90^0$ directions from the laser beam axis. We have considered the plasma generated by the laser beam in water as an antenna for acoustic signal transmission in our methods. Therefore, the localization message block containing the GPS coordinates of the plasma is sent to the UWN from air by focusing laser beam in water. The UWN uses Received Signal Strength (RSS) to measure the distances from acoustic transmitters and consequently estimates its own GPS coordinates. We devise two localization techniques, one for static and another for dynamic UWNs. Finally, we evaluate the effectiveness of our approach through simulation using our experimentally measured data. The simulation results show that both techniques can achieve the same accuracy as traditional methods without surface or underwater anchor nodes.

The paper is organized as follows. Section II covers related work. The optoacoustic signal generation is analyzed in Section III. The RSS-based ranging and the acoustic propagation model are discussed in Section IV. Section V describes our proposed localization techniques using optoacoustic signals. Section VI reports the performances and discusses the localization errors. The paper is concluded in Section VII.

## II. RELATED WORK

In recent years, many methods have been proposed for underwater node localization. Since acoustic communication is the prime choice for underwater environments, most of the published localization techniques exploit acoustic signals [7]. However, these techniques require the deployment of multiple floating gateway nodes on the water surface [8][9], or tethered anchor nodes with known underwater positions [10], or both [11]. Another method is to deploy mobile anchor nodes that receive GPS coordinates when on the water surface and periodically propagate its current location information while traveling underwater [1][12]. Recently, a localization scheme has been proposed to eliminate the inter-medium gateway node using visible light communications (VLC) [13]. However, the range of cross-medium VLC transmissions is only a few meters.

This paper focuses on developing a localization method for dispersed UWNs that eliminates the need for surface gateways and anchor nodes. We leverage the advantages of cross-medium optoacoustic communication which has a longer underwater reach. The optoacoustic signals will be modulated to convoy the global coordinates of the incident points on the surface. The optoacoustic process has been widely studied in the literature. Vogel et al. [14] have characterized the shock wave and bubble generation; the effects of various laser repetition rates are studied in [15]. We have also devised a novel modulation technique for optoacoustic signals [4]. Moreover, the acoustic signal generated from the optoacoustic process can generally propagate further than visible light in underwater setups. For example, the underwater wireless optical signal can travel typically less than 100 m [16], while the generated acoustic pulse propagation was measured at distances up to 300 m for an acoustic source level (SL) of about 190 dB re µPa at 1 m in [17]. Thus, we can achieve even more localization range for higher SL like 210 dB re µPa at 1 m reported in [6]. To the best of our knowledge, no prior work has pursued UWN localization using the optoacoustic transmissions from airborne units.

## III. OPTOACOUSTIC SIGNAL GENERATION

Acoustic signal is generated when high intensity light impinges on a liquid medium like water. For example, we can generate an optoacoustic signal by focusing a high energy laser beam on a small spot in water and create an optical breakdown. Laser-induced optical breakdown is a nonlinear absorption process in which the breakdown threshold irradiance is exceeded, resulting in plasma generation. This plasma formation is associated with breakdown shockwave, cavitation bubble oscillation-produced shock waves. The pulse duration of the laser affects the breakdown threshold. A. Vogel et al. [18] have investigated the thresholds for various pulse durations and focusing angles combinations. For a few nanosecond pulse durations, the irradiance threshold values are in the order of $10^{11}$ W/cm$^2$ and $10^{13}$ W/cm$^2$ for 100 femtosecond pulse duration in order to generate plasma in water [18]. Fig. 1 shows the shock wave generated from the optical breakdown in water for different laser parameters. This figure is regenerated from [14] where the authors have studied the shock wave emission and cavitation bubble expansion with 30 ps and 6 ns Nd:YAG laser pulses for energies between 50 $\mu J$ and 10 mJ. We can observe that the shock wave velocity and duration vary with different laser parameters and increase when laser pulse energy increases. The initial velocity of the shock wave is very high but quickly reduces to the sound velocity in water.

In our proposed localization method, we will send the coordinates information from a remote position in the air as an alternative to traditional approach for generating underwater acoustic signals that rely on the use of submerged transducers. Laser-induced underwater plasma is considered as the antenna and the volume and shape of this plasma are important because they determine the duration and directivity of the generated acoustic pulses. Generally, non-spherical shaped plasma generates anisotropic acoustic pressure. However for the UWN localization, we need the same acoustic pressure in all the directions for accurate distance measurement. Thus, more spherical shaped plasma is needed which can generate isotropic pressure. The shape of the plasma can be changed by varying the focusing angle of the laser [4]. In order to get the same pressure in all directions, we can vary the laser focusing angle. The dependency of maximum plasma length ($z_{max}$) on the focusing angle ($\theta$) of the lens is given in [19] as,

$$z_{max} = \frac{\lambda}{\pi \, tan^2 \frac{\theta}{2}} \sqrt{\beta - 1} \qquad (1)$$

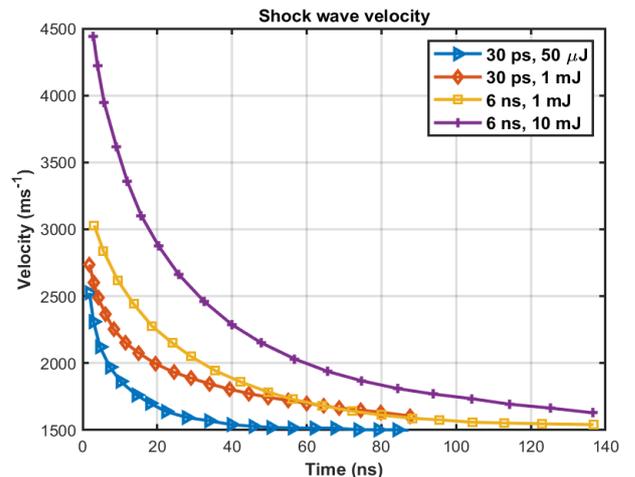

**Figure 1:** Experimentally determined shockwave velocity [14].

Here, the normalized laser pulse energy, $\beta = \frac{E}{E_{th}} = \frac{I}{I_{th}}$, $E$ and $E_{th}$ are the laser pulse energy and breakdown threshold energy, respectively, and $\lambda$ is the wavelength of the laser beam. It is evident from (1) that the plasma will be more elongated for higher energy laser pulses. Moreover, $z_{max}$ is dependent on the focusing angle and focal spot radius, which are inversely related. Thus, increasing the focusing angle will decrease the spot size, and consequently the plasma length will decrease.

A laboratory experiment has been conducted to demonstrate nonlinear optoacoustic signal generation. The experimental arrangement for generating the same acoustic pressure is depicted in Fig. 2 (a). We have used a Nd:YAG laser emitting 6 ns pulses at a wavelength of 1064 nm. The laser pulses are focused on water with a convex lens. Absorption material was placed on all the sides of the water tank to absorb sound reflections. In order to investigate the acoustic signal generated in all the directions, we have taken the acoustic signal data in three steps. In these steps, we placed the hydrophone in the $0^0$, $45^0$ and $90^0$ directions with respect to the laser beam axis. Laser pulse energy and the focusing lens were varied to get the same acoustic peak pressure in all three directions. Fig. 2 (b) shows the peak-peak voltage measured by the hydrophone in $0^0$, $45^0$ and $90^0$ directions with a 30 mJ laser pulse focused with a 75 mm lens. The peak-peak voltage has varied the most in the $0^0$ direction, and the least in $90^0$ direction. However, the mean value is almost the same in all directions.

## IV. RSS-Based Distance Measurement

The accuracy of any localization technique depends on the precision of the underlying ranging measurements. Popular ranging methods include Time of Arrival (ToA), Time Difference of Arrival (TDoA), Received Signal Strength (RSS), and Angle of Arrival (AoA). Although ToA-based localization is widely used, it requires precise clock synchronization for the communicating nodes, which is very difficult, if not even impossible, to achieve between aerial and underwater nodes. In a ToA-based distance measurement method, the speed of underwater sound is multiplied by the time difference between transmitted and received signals. In the optoacoustic process, optical breakdown generated shock wave velocity is way higher; from Fig. 1, we can observe that shock wave velocity and duration vary with different laser parameters. In addition, measuring shockwave velocity and duration is very challenging, which hinders accurate distance measurement using ToA. In this optoacoustic process, the laser beam travels from the air and focuses underwater. Due to the differences in mediums, the laser propagation speed varies as well across the mediums and also not the same with acoustic signal's velocity. Therefore, ToA-based ranging is deemed impractical in our context. The TDoA method, on the other hand, needs multiple airborne nodes. Meanwhile, AoA is prone to high errors given the variability of the plasma shape. Thus, we are considering the RSS-based distance measurement.

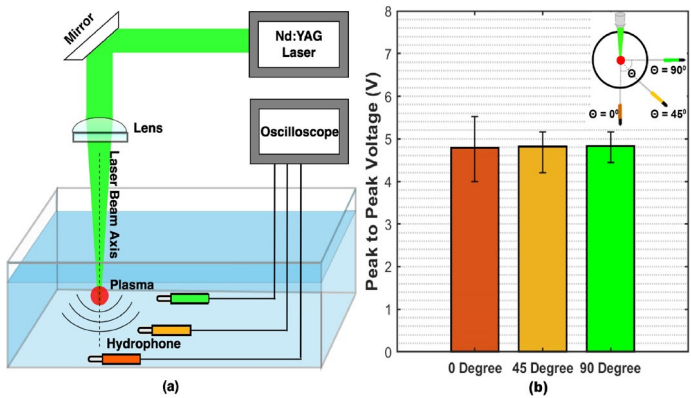

**Figure 2:** (a) Experimental setup for generating optoacoustic signals; (b) Peak-peak voltage generated from acoustic signals in $0^0$, $45^0$ and $90^0$ directions. Each presented value is the mean of ten measurements with the error bar showing the maximum and minimum values.

| Control Bits | SL | Coordinates |

**Figure 3:** Localization message block.

The aerial node can determine its GPS coordinates and use the lens's focal length to calculate the plasma location. In our model, the underwater plasma radiates acoustic signal isotropically. The UWN utilizes measured RSS of the emitted acoustic signal from the plasma to estimate proximity. The localization message block is shown in Fig. 3. First, control bits are sent to the unlocalized UWN to calibrate and calculate the mean sound intensity level (SIL). The experiments show that the acoustic signal's generated SL is not precisely the same for every laser pulse, and therefore, the received signal's SIL can be slightly different for multiple measurements. Thus, taking a mean of the control bits SIL values to calculate the acoustic signal propagation's transmission loss (TL) would be more accurate. Initial signal strength (SL) is assumed to be known by the aerial node for its particular laser parameters and sent within the same packet to the unlocalized UWN, after the control bits field as shown in Fig. 3. Finally, the coordinates of the acoustic transmitter, which is the plasma location, are sent to the UWN. After receiving the localization message block, the receiver UWN can calculate the TL using,

$$TL = SL - SIL \quad (2)$$

Attenuation of the acoustic signal is frequency dependent and is proportional to the distance between the plasma and the UWN receiver. The acoustic signal propagation is weakened in the ocean primarily due to two phenomena, namely, spreading and absorption [20]. The total TL is given by,

$$TL = 10 \cdot k \log d + (\alpha \cdot D \cdot 10^{-3}) \quad (3)$$

Where $k$, $D$ and $\alpha$ are the spreading factor, the distance from plasma to the receiver UWN, and the absorption coefficient, respectively. In (3), the first part is for spreading loss and the second part reflects absorption loss. The spreading loss is a result of acoustic wave geometric propagation away from the source. Cylindrical and spherical spreading are two simple approximations used to describe the spreading loss. The usual values of $k$ are 1, 1.5 and 2 for cylindrical, practical and spherical spreading, respectively. We are considering spherical spreading in our model. When sound propagates across the ocean, a portion of its acoustic energy is continuously absorbed and converted into heat. This absorption is caused mainly by the liquid's viscosity, particularly at frequencies between 100 Hz and 100 kHz [21]. Another factor for the decrease in sound intensity with distance in the water is the scattering of sound waves caused by numerous types of inhomogeneities. Usually, the combined effect of absorption and scattering can only be quantified. The absorption coefficient ($\alpha$) in dB/km for frequency $f$ in kHz is obtained from Thorp's formula [22],

$$\alpha = \frac{0.11 f^2}{1 + f^2} + \frac{44 f^2}{4100 + f^2} + 2.75 \cdot 10^{-4} f^2 + 0.003 \quad (4)$$

Equation (4) is generally used for frequencies between 100 Hz to 3 kHz. Using Schulkin and March model [23] the $\alpha$ for the frequency range between 3 kHz and 500 kHz can be calculated by,

$$\alpha = 8.68 \cdot 10^3 \left(\frac{SAf_Tf^2}{f_T^2 + f^2} + \frac{Bf^2}{f_T}\right)(1 - 6.54 \cdot 10^{-4}P) \quad (5)$$

Where $A = 2.34 \times 10^{-6}$ and $B = 3.38 \times 10^{-6}$ are constants, $S$ (‰) is the salinity, $P$ (kg/cm2) is the hydrostatic pressure, $f$ (kHz) is the acoustic wave frequency. The relaxation frequency $f_T$ (kHz) is expressed by,

$$f_T = 21.9 \cdot 10^{6 - 1520/(T+273)} \quad (6)$$

Here, $T$ (°C) is the temperature of the water. Hosseini et al. [24] invert the TL from (3) using the Lambert $W$ function and calculate $D$ using the Halley method,

$$D = \frac{20000 \times W\left(\left(\frac{\ln(10)}{20000}\right)\alpha e^{(\ln(10)/20)\times TL}\right)}{\alpha \times \ln(10)} \quad (7)$$

Thus, the UWN can measure its distance from the plasma using (7). This is an iterative process and using the Lambert $W$ function is more efficient than Newton-Raphson inversion and capable of calculating accurate distance as fast as four iterations.

## V. UWN Localization Using Optoacoustic Signals

### A. Static Underwater Node

We consider the typical underwater environment shown in Fig. 4, and employ an airborne node to transmit localization messages to the UWNs with unknown positions by focusing a laser beam in water. The plasma generated underwater by the laser will act as an acoustic transmitter. In our system, the airborne node is equipped with a GPS receiver and uses the focal length of the focusing lens to calculate the coordinates of the plasma. The UWN is assumed to have a pressure sensor to calculate its depth from the water surface. If the UWN is static during the localization process, the airborne node needs to move to at least three noncollinear positions to transmit the localization message block to the UWN. As a result, the UWN should have at least three reference points with GPS coordinates along with the localization message block and consequently estimate its position using multilateration.

Multilateration is the most common method for determining a position using proximity to reference points. For example, it is assumed that the airborne node moves to $n$ different noncollinear positions and transmits the localization message block to the UWN for localization. If the coordinate of the $i^{th}$ position of the plasma is $(x_i, y_i, z_i)$, the UWN can estimate the distance from the plasma with the below expression,

$$(x - x_i)^2 + (y - y_i)^2 + (z - z_i)^2 = D_i^2 \quad (8)$$

where the coordinates of the target UWN are $(x, y, z)$. Thus, the UWN can derive $n$ equations from $n$ different positions of the airborne node. The system can be linearized by subtracting the last equation from the first $n - 1$ equations and can be expressed as $A\varphi = b$ where,

$$A = \begin{bmatrix} 2(x_1 - x_n) & 2(y_1 - y_n) \\ \vdots & \vdots \\ 2(x_{n-1} - x_n) & 2(y_{n-1} - y_n) \end{bmatrix} \quad (9)$$

and,

$$b = \begin{bmatrix} x_1^2 - x_n^2 + y_1^2 - y_n^2 + z_1^2 - z_n^2 \\ -2z(z_1 - z_n) + d_n^2 - d_1^2 \\ \vdots \\ x_{n-1}^2 - x_n^2 + y_{n-1}^2 - y_n^2 + z_{n-1}^2 - z_n^2 \\ -2z(z_{n-1} - z_n) + d_n^2 - d_{n-1}^2 \end{bmatrix} \quad (10)$$

Here, $z$ coordinate of the UWN is present in matrix $b$ because UWN can measure it using the pressure sensor. The UWN's unknown coordinates are $\hat{\varphi} = [\hat{x}\hat{y}]^T$ and we can find them with the least square method,

$$\hat{\varphi} = (A^TA)^{-1}A^Tb \quad (11)$$

Fig. 4 illustrates a minimum scale scenario where a single airborne node moves to only three noncollinear positions and each of the deployed UWN ($U_a$, $U_b$, $U_c$) receives three localization message blocks. In deeded the airborne node can mover to more positions to provide more reference points and increase the UWN localization accuracy.

### B. Dynamic Underwater Node

This technique is proposed for dynamic UWN which is also assumed to be equipped with a pressure sensor to measure the depth from the water surface. Here, only one position of the airborne node is required to focus the laser beam into the water to transmit the localization message block. The unknown positioned dynamic UWN receives the localization message block from a certain position, then moves to two different positions in the x-y plane and receives two more localization message blocks. Fig. 5 depicts the movement of the UWN, where it receives the first localization message block at point "A", then moves to point "B" and finally to point "C". The coordinates of A, B, and C and the plasma (acoustic source) $S$ are denoted by $(x_a, y_a, z_a)$, $(x_b, y_a, z_a)$, $(x_b, y_c, z_a)$ and $(x_s, y_s, z_s)$, respectively. All three points have the same $z$ coordinate because the UWN moves only in the x-y plane and $z_a$ is determined by the UWN using its pressure sensor. In this technique, the UWN moves to point $B$ by varying only $x$ coordinate and then to $C$ by varying only the $y$ coordinate. Thus, the relationship between coordinates can be expressed by,

$$x_b = x_a + D_{AB} \quad (12)$$

$$y_c = y_a + D_{BC} \quad (13)$$

Where $D_{AB}$ and $D_{BC}$ are the distance between points $A$ and $B$, and between $B$ and $C$, respectively. It is assumed that the UWN can measure the distance $D_{AB}$ and $D_{BC}$, e.g. using speedometer. Therefore, the UWN can derive three equations using (8) for points A, B and C and use (12) and (13) to find the unknown coordinates of the UWN's final position $C$ as,

$$x_b = \frac{D_B^2 - D_A^2 + D_{AB}^2 + 2D_{AB}x_s}{2D_{AB}} \quad (14)$$

$$y_c = \frac{D_C^2 - D_B^2 + D_{BC}^2 + 2D_{BC}y_s}{2D_{BC}} \quad (15)$$

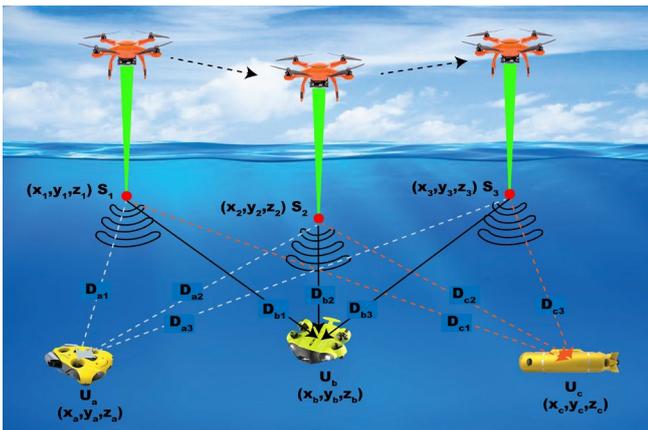

**Figure 4:** Static underwater node localization by airborne node using optoacoustic signals.

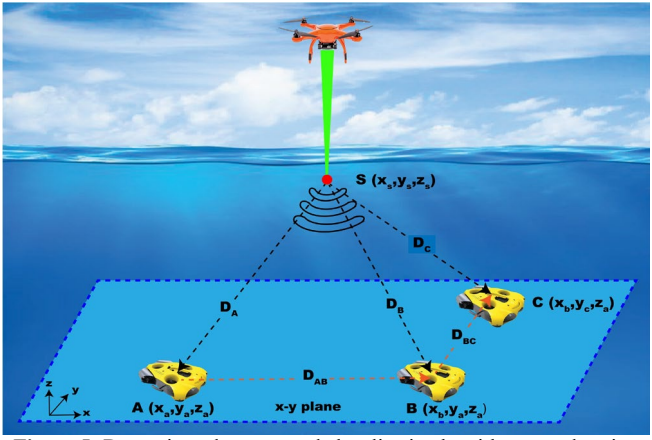

**Figure 5:** Dynamic underwater node localization by airborne node using optoacoustic signals.

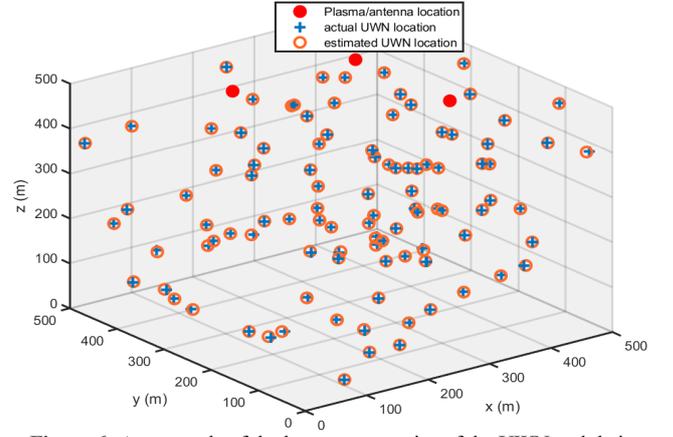

**Figure 6:** An example of deployment scenarios of the UWN and their estimated locations.

Where, $D_A$, $D_B$ and $D_C$ are the distances of the UWN position $A$, $B$ and $C$ from the plasma, i.e., point $S$, respectively. The UWN can calculate $D_A$, $D_B$ and $D_C$ using (7). Thus, the unknown positioned UWN can be localized by receiving localization message blocks corresponding to three different positions of the airborne node.

## VI. Performance Analysis

We have used Matlab simulation to validate our proposed underwater localization methods and compared the results with traditional technique based on surface or underwater anchor based localization. The simulation models optoacoustic communications based on empirical measurements from experiments conducted in our lab, where the generated acoustic signals are measured as different directions relative to the laser beam axis, , e.g. $0^0$, $45^0$ and $90^0$. The peak-peak voltage measured by the hydrophone is used as the control bits and the mean value of the peak-peak voltage is used as the SL in our simulations.

Our simulation setup for static UWN localization can be summarized in Fig. 6, where plasma is created in three noncollinear locations for transmitting the localization message blocks. Additive White Gaussian Noise (AWGN) channel is considered for acoustic signal transmission underwater. In our laboratory experiments, we have found the $f_{peak}$ (the frequency with the highest power within the distribution of the acoustic pulses component frequencies) of the acoustic signals is around 8 kHz. Thus we have used the Schulkin and March model for $\alpha$ calculation. In the simulation settings, 100 unknown positioned UWN is placed randomly in a three-dimensional 500×500×500 cubic meters area and using our method, the UWN location is estimated. The accuracy of our method is measured by The Root Mean Square Error (RMSE) and compared with the traditional technique where the isotropic acoustic signal is generated using submerged acoustic transducers. The RMSE is expressed by,

$$RMSE = \sqrt{\frac{\sum_{i=1}^{N}(x_i - \hat{x}_i)^2 + (y_i - \hat{y}_i)^2 + (z_i - \hat{z}_i)^2}{N}} \quad (16)$$

Here, the actual UWN location is $(x, y, z)$, the estimated location is $(\hat{x}, \hat{y}, \hat{z})$ and $N$ is the total number of unknown UWN. In Fig. 6, the UWN position is estimated with Signal to Noise (SNR) value of 30 dB.

Figures 7, 8 and 9 show the RMSE for the static UWN setting, where the UWN is located around $0^0$, $45^0$ and $90^0$, relative to the laser beam (norm on the water surface), respectively. The results for our method is compared with the baseline approach, i.e., techniques that deploy surface or underwater anchors . We have varied the number of control bits in the localization message block and found that for only 16 control bits, our method achieves the same accuracy as the baseline, as shown in Figures 7 and 8. Such accuracy is affected

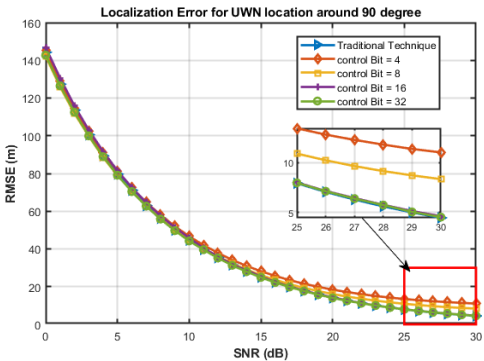

**Figure 7:** RMSE vs SNR for UWN location around $90^0$ direction from the laser beam axis.

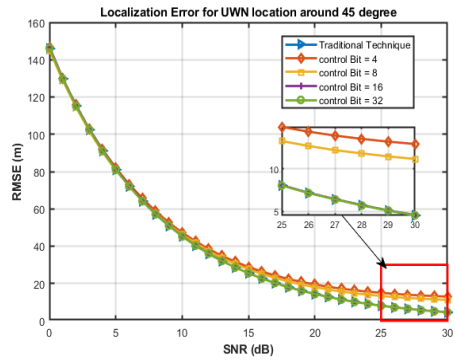

**Figure 8:** RMSE vs SNR for UWN location around $45^0$ direction from the laser beam axis.

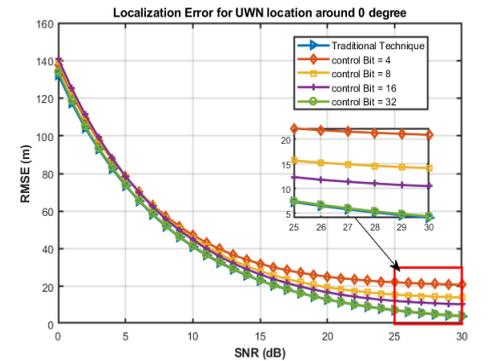

**Figure 9:** RMSE vs SNR for UWN location around $0^0$ direction from the laser beam axis.

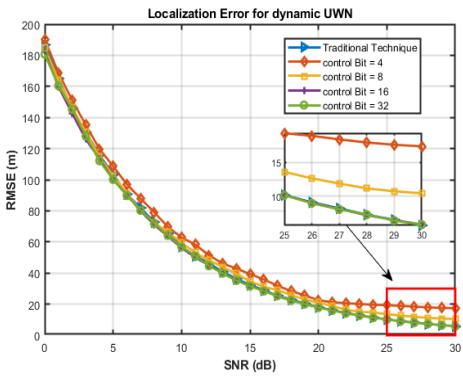
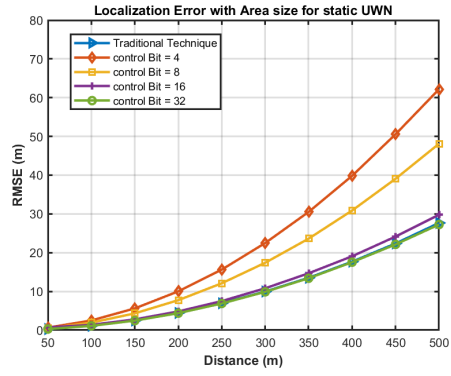
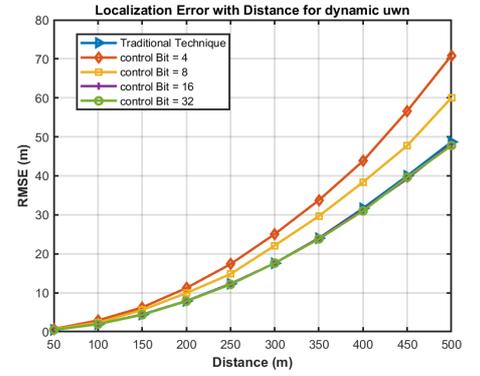

**Figure 10:** RMSE vs SNR for dynamic UWN.    **Figure 11:** RMSE with area size for static UWN.    **Figure 12:** RMSE vs. area size for dynamic UWN.

by the node position relative to the laser beam, where Fig. 9 indicates that 32 control bits would be needed for our method to match the accuracy of the the baseline approach because the experimentally generated acoustic signal varies the most in the $0^0$ direction. Overall, in all three locations of the static UWN, the RMSE is below 5 m for 30 dB SNR.

We used a similar simulation setup for the dynamic UWN localization technique and randomly varied the UWN location by changing only x coordinates and then y coordinates. Fig. 10 shows the RMSE for such a dynamic UWN localization setup. We can observe that the RMSE of this case is higher than the static UWN and achieves the same accuracy as the traditional technique for 16 control bits. In Figures 11 and 12, the distance between the plasma and UWN is varied to capture the effect on RMSE for the static and dynamic UWN cases, respectively. We can observe that the localization error is higher for longer distances from the acoustic source, which is expected. Such an error is less significant for the static UWN setup, because the position of the dynamic UWN is changed randomly by changing only one coordinate. Therefore, sometimes the three positions are not noncollinear enough to give the precise location.

## VII. CONCLUSIONS

This paper has presented a novel methodology for localizing and providing GPS coordinates to underwater nodes using optoacoustic signals. We have analyzed the process of controlling the shape and size of the plasma to create the isotropic acoustic transmitter and experimentally validate its generation. We have devised two techniques for handling static and dynamic underwater nodes. The effectiveness of our approach has been confirmed through simulation and compared with traditional techniques where submerged acoustic transducers are used. The validation results have shown that our method can achieve the same accuracy as traditional techniques without using any surface or underwater anchor nodes.

Acknowledgement: This work is supported in part by the National Science Foundation, USA, contract #0000010465.

## REFERENCES

[1] M. Erol, L. F. M. Vieira, and M. Gerla, "Localization with Dive'N'Rise (DNR) beacons for underwater acoustic sensor networks," *Proc. of the 2nd workshop on Underwater networks (WuWNet)*, Montreal, Quebec, Canada, 2007, pp. 97–100.

[2] M. S. Islam and M. F. Younis, "Analyzing visible light communication through air–water interface," *IEEE Access*, 7, pp. 123830-123845, 2019.

[3] D. Pompili and I. F. Akyildiz, "Overview of Networking Protocols for Underwater Wireless Communications," *IEEE Communications Magazine*, Vol. 47, No. 1, pp. 97 – 102, Feb. 2009.

[4] M. Mahmud, M. S. Islam, M. Younis and G. Carter, " Optical Focusing-based Adaptive Modulation for Optoacoustic Communication," *Proc. of 30th Wireless and Optical Comm. Conf. (WOCC)*, Taipei, Taiwan, 2021.

[5] Blackmon, F, Estes, L., and Fain, G., "Linear opto-acoustic underwater communication," *Applied Optics*, Vol. 44, No. 18, June, 2005

[6] T. G. Jones, M. Hornstein, A. Ting, D. Gordon and Z. Wilkes, "Characterization of underwater laser acoustic source for navy applications," *Proc. of the IEEE International Conference on Plasma Science* - Abstracts, San Diego, CA, 2009.

[7] J. Luo, Y. Yang, Z. Wang and Y. Chen, "Localization Algorithm for Underwater Sensor Network: A Review," *IEEE Internet of Things Journal*, vol. 8, no. 17, pp. 13126-13144, 1 Sept.1, 2021.

[8] T. L. N. Nguyen and Y. Shin, "An efficient RSS localization for underwater wireless sensor networks," *Sensors*, vol. 19, no. 14, p. 3105, 2019.

[9] F. Liu, H. Chen, L. Zhang and L. Xie, "Time-Difference-of-Arrival-Based Localization Methods of Underwater Mobile Nodes Using Multiple Surface Beacons," *IEEE Access*, vol. 9, pp. 31712-31725, 2021.

[10] C. Zheng, D. Sun, L. Cai and X. Li, "Mobile Node Localization in Underwater Wireless Networks," *IEEE Access*, vol. 6, pp. 17232-17244, 2018.

[11] T. Islam and Y. K. Lee, "A two-stage localization scheme with partition handling for data tagging in underwater acoustic sensor networks," *Sensors (Basel)*, vol. 19, no. 9, p. 2135, 2019.

[12] J. Luo, Y. Yang, Z. Wang, Y. Chen, and M. Wu, "A mobility-assisted localization algorithm for three-dimensional large-scale UWSNs," *Sensors (Basel)*, vol. 20, no. 15, p. 4293, 2020.

[13] J. Bin Saif, and M. Younis, "Underwater Localization using Airborne Visible Light Communication Links," *Proc. the IEEE Global Comm. Conf. (GLOBECOM 2021)*, Madrid, Spain, December 2021 (to appear).

[14] A. Vogel, S. Busch, and U. Parlitz, "Shock wave emission and cavitation bubble generation by picosecond and nanosecond optical breakdown in water," *J. Acoust. Soc. Am.*, vol. 100, no. 1, pp. 148–165, 1996.

[15] F. Blackmon and L. Antonelli, "Experimental demonstration of multiple pulse nonlinear optoacoustic signal generation and control," *Appl. Opt.*, vol. 44, no. 1, pp. 103–112, 2005.

[16] M. V. Jamali, A. Chizari, and J. A. Salehi, "Performance Analysis of Multi-Hop Underwater Wireless Optical Communication Systems," *IEEE Photonics Technology Letters*, 29, pp. 462-465, Jan. 2017.

[17] T.G. Jones, M. Helle, A. Ting, and M. Nicholas, "Tailoring Underwater Laser Acoustic Pulses," *NRL REVIEW, acoustics*, pp. 142-143, 2012.

[18] A. Vogel, J. Noack, G. Hüttman, and G. Paltauf, "Mechanisms of femtosecond laser nanosurgery of cells and tissues," *Appl. Phys. B*, vol. 81, no. 8, pp. 1015–1047, 2005.

[19] A. Vogel, K. Nahen, D. Theisen, and J. Noack, "Plasma formation in water by picosecond and nanosecond Nd:YAG laser pulses. I. Optical breakdown at threshold and superthreshold irradiance," *IEEE Journal of Selected Topics in Quantum Electronics*, vol. 2, no. 4, pp. 847–860, 1996.

[20] Leonid Maksimovich Brekhovskikh, Yu. P. Lysanov, "Fundamentals of ocean acoustics, Third Edition," *Springer-Verlag*, NY, 2003.

[21] F. De Rango, F. Veltri, and P. Fazio, "A multipath fading channel model for underwater shallow acoustic communications," in Proc. IEEE Int. Conf. Commun., Ottawa, ON, Canada, 2012, pp. 3811—3815.

[22] W. H. Thorp, "Analytic description of the low-frequency attenuation coefficient," *J. Acoust. Soc. Am.*, vol. 42, no. 1, pp. 270–270, 1967.

[23] H. W. Marsh and M. Schulkin, "Report on the status of Project AMOS Tch," U.S.Navy Underwater Sound Lab., New London, CT, 1952.

[24] M. Hosseini, H. Chizari, T. Poston, M. B. Salleh, and A. H. Abdullah, "Efficient underwater RSS value to distance inversion using the Lambert function," *Math. Probl. Eng.*, vol. 2014, pp. 1–8, 2014.